\documentclass[3p,times,procedia]{elsarticle}
\usepackage{settings}

\newcommand\CSC{\textsc{csc}}
\newcommand\doe{\textsc{doe}}

\newcommand\lhc{\textsc{lhc}}
\newcommand\moe{\textsc{moe}}

\newcommand\nsf{\textsc{nsf}}
\newcommand\nsfc{\textsc{nsfc}}
\newcommand\rhic{\textsc{rhic}}

\newcommand\alice{\textsc{alice}}

\newcommand\cms{\textsc{cms}}
\newcommand\phenix{\textsc{phenix}}

\newcommand\STAR{\textsc{star}}

\newcommand\clvisc{\textsc{clv}isc}

\newcommand\jetscape{\textsc{jetscape}}

\newcommand\qcd{\textsc{qcd}}

\newcommand\qgp{\textsc{qgp}}

\let\abs\undefined\DeclarePairedDelimiter\abs{\lvert}{\rvert}

\newcommand\dd{\mathop{}\!\romanup{d}}

\DeclareSIUnit{\fm}{\femto\metre}

\newcommand\proton{{\romanup{p}}}

\newcommand\Bmeson{{\romanup{B}}}

\newcommand\Dmeson{{\romanup{D}}}

\newcommand\A{{\romanup{A}}}

\newcommand\nAu{{\romanup{Au}}}
\newcommand\nPb{{\romanup{Pb}}}

\newcommand\nuclei{{\A\A}}
\newcommand\pp{{\proton\proton}}

\newcommand\AuAu{{\nAu\text{+}\nAu}}
\newcommand\PbPb{{\nPb\text{+}\nPb}}

\newcommand\snn{\sqrt{s_\text{NN}}}
\newcommand\snnG[1]{\snn = \SI{#1}{\GeV}}
\newcommand\snnT[1]{\snn = \SI{#1}{\TeV}}

\newcommand\pt{p_\text{T}}

\newcommand\raa{R_\text{AA}}

\renewcommand\textsc{\MakeUppercase}

\biboptions{sort&compress,elide}

\usepackage{nupha_ecrc}
\journalname{Nuclear Physics A}
\volume{00}
\firstpage{1}
\runauth{Caio A.~G.~Prado et al.}
\jid{nupha}
\jnltitlelogo{Nuclear Physics A}

\begin{document}
\begin{frontmatter}
    \title     {Longitudinal dependence of $B$ and $D$ meson nuclear modifications in heavy-ion collisions at \rhic\ and the \lhc}

    \newcommand\fixmarks{$^{\text{a,}\ast}$}  
    \author                    {Caio A.~G.~Prado\fixmarks}
    \author [add:moe]          {Wen-Jing Xing}
    \author [add:wsu,add:tamu] {Shanshan Cao} 
    \author [add:moe,add:bnl]  {Guang-You Qin}
    \author [add:moe,add:bnl]  {Xin-Nian Wang}
    \cortext[]{\parbox[t]{\linewidth}
        {Presenter of the parallel talk\\
        \emph{Email address:} \texttt{cagprado@mail.ccnu.edu.cn}}}

    \address[add:moe]  {Institute of Particle Physics and Key Laboratory of Quark and Lepton Physics (\moe), Central China Normal University, Wuhan, Hubei, 430079, China}
    \address[add:wsu]  {Department of Physics and Astronomy, Wayne State University, Detroit, MI 48201, USA}
    \address[add:tamu] {Cyclotron Institute, Texas A\&M University, College Station, TX 77843, USA}
    \address[add:bnl]  {Nuclear Science Division, Lawrence Berkeley National Laboratory, Berkeley, CA 94720, USA}

    \dochead{XXVIIIth International Conference on Ultrarelativistic Nucleus-Nucleus Collisions\\ (Quark Matter 2019)}

    \begin{abstract}
        It is widely acknowledged that heavy flavor probes are sensitive to the properties of the quark-gluon plasma and are often considered an important tool for the plasma tomography studies.
        Forward rapidity observables can provide further insight on the dynamics of the medium due to the interplay between the medium size and the differences in the production spectra of heavy quark probes.
        In this proceedings we present the nuclear modification factor $\raa$'s for $\Bmeson$ and $\Dmeson$ mesons, as well as heavy flavor leptons, in the rapidity range $-4.0 < y < 4.0$ obtained from relativistic Langevin equation with gluon radiation coupled with a (3+1)-dimensional viscous hydrodynamics medium background.
        We present comparison with experimental data at mid-rapidity as well as predictions for different rapidity ranges.
    \end{abstract}

\end{frontmatter}
\section{Introduction}

Heavy quarks are extremely valuable probes for the tomographic study of the hot and dense nuclear matter known as the quark-gluon plasma (\qgp) produced in relativistic heavy-ion collisions~\cite{Wang:1991xy,Qin:2015srf,Moore:2004tg,Borsanyi:2010bp,Dong:2019byy}.
Final observables from heavy quarks, such as the nuclear modification factor, contain cumulative information of the evolution dynamics of the \qgp.

One of the most common observables pertaining heavy flavor studies is the nuclear modification factor $\raa$, which is usually associated with parton energy loss through the medium.
It is defined as the ratio between the particle spectrum in nuclei collisions $\dd N_{\nuclei}/\dd \pt$, and the spectrum in $\pp$ collisions, $\dd N_{\pp}/\dd \pt$~\cite{Miller:2007ri}:
\begin{equation}\label{eq:raa}
\raa(\pt,y) = \frac{1}{\mathcal{N}}\frac{\dd N_{\nuclei}/\dd \pt \dd y}{\dd N_{\pp}/\dd \pt \dd y},
\end{equation}
where $\mathcal{N}$ is the average number of binary nucleon-nucleon collisions for a given centrality class of $\nuclei$ collisions.

Many studies attempted to use $\raa$ to investigate mechanisms of parton transport and energy loss or to constraint phenomenological models together with other observables.
However, it was not until recently that longitudinal dependence of heavy flavor observables started to be explored~\cite{Das:2016cwd,Chatterjee:2017ahy,Nasim:2018hyw,Zhang:2019hzn,Adam:2019wnk}.
In this direction we recently calculated the nuclear modification factor $\raa$ of open heavy flavor in a wide range of rapidity~\cite{Prado:2019ste}.

In this talk we present the longitudinal dependence of the $\raa$ of heavy flavor mesons ($\Bmeson$ and $\Dmeson$) as well as electrons and muons decayed from these particles~\cite{Prado:2019ste}.
We use the three dimensional medium profiles generated from \clvisc\ hydrodynamics code~\cite{Pang:2018zzo,Pang:2012he,Wu:2018cpc} to construct averaged \qgp\ backgrounds in which heavy quarks are sampled and allowed to propagate using relativistic Langevin equations with gluon radiation and a hybrid fragmentation plus coalescence model for hadronization~\cite{Cao:2013ita,Cao:2015hia}.
The initial position of the sampled heavy quarks is obtained from Monte Carlo Glauber model while their initial momentum distribution is calculated using leading order perturbative \qcd~\cite{Combridge:1978kx} that includes flavor excitation, pair production, and nuclear shadowing and anti-shadowing effects~\cite{Cao:2015hia,Lai:1999wy,Eskola:2009uj}.
The initial production cross section is also used to calculate the relative fraction between bottom and charm quarks in order to obtain heavy quark decayed leptons spectra.
Predictions are made for different rapidity bins in the range of $-4.0 < y < 4.0$ on the nuclear modification factor of open heavy flavor mesons.

\section{Numerical Results}

\begin{figure}[b!]
    \includegraphics{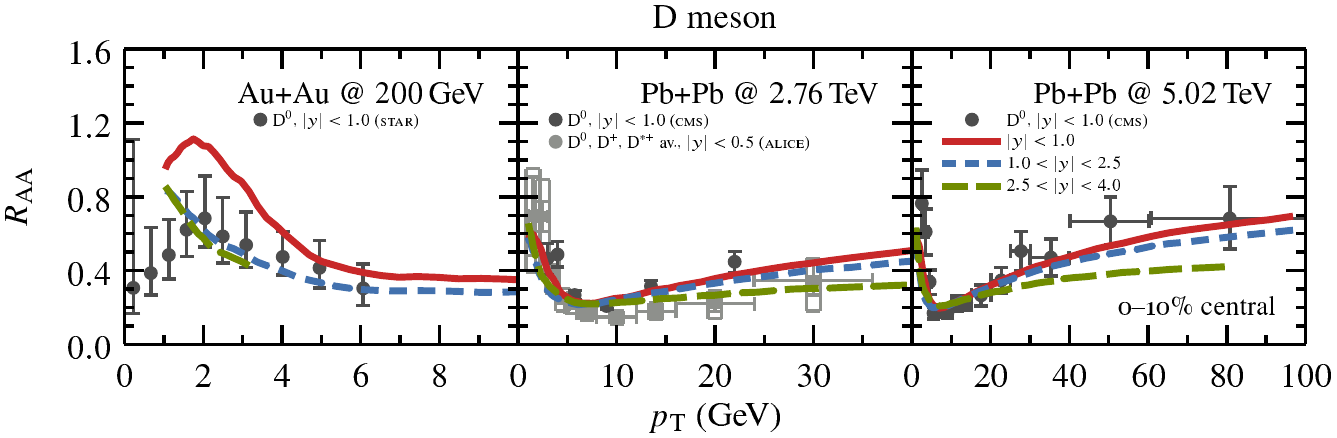}
    \caption{(Color online) Nuclear modification factor of $\Dmeson$ mesons for central collisions in different ranges of rapidity.  Mid-rapidity experimental data from \STAR~\cite{Radhakrishnan:2019gbl} (left), \alice~\cite{Adam:2015sza} and \cms~\cite{CMS:2015hca} (middle), and \cms~\cite{Sirunyan:2017xss} (right) are also shown.}
    \label{fig:raaD}
\end{figure}

\begin{figure}[t!]
    \includegraphics{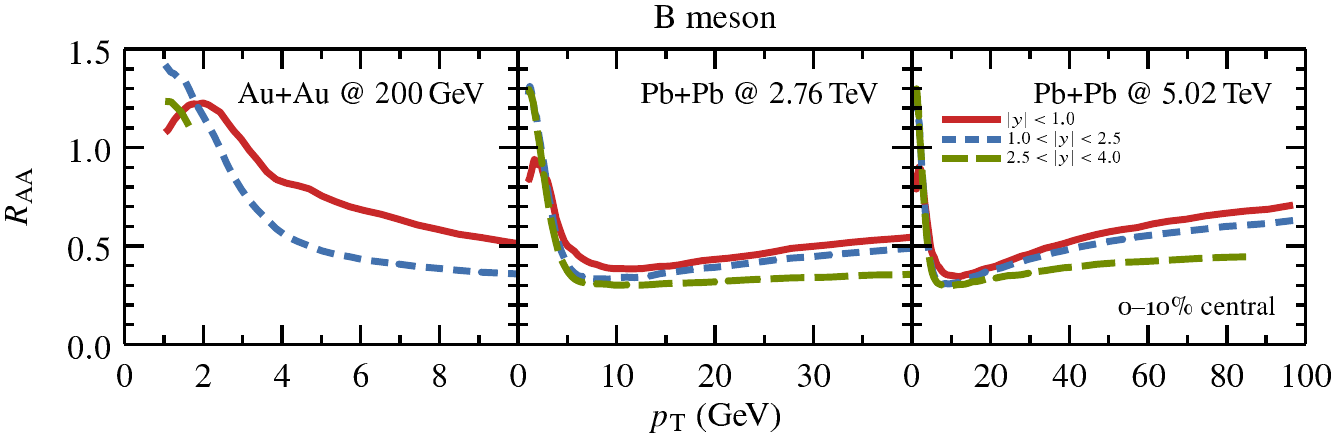}
    \caption{(Color online) Nuclear modification factor of $\Bmeson$ mesons for central collisions in different ranges of rapidity for $\AuAu$ at $\snnG{200}$ (left), $\PbPb$ at $\snnT{2.76}$ (middle) and $\PbPb$ at $\snnT{5.02}$ (right).}
    \label{fig:raaB}
\end{figure}

Using the setup as described in reference~\cite{Prado:2019ste}, we present the numerical results from our simulations.
Calculations for $\Dmeson$ meson $\raa$ are shown in Fig.~\ref{fig:raaD} for collisions of $\AuAu$ at $\snnG{200}$, $\PbPb$ at $\snnT{2.76}$, and $\PbPb$ at $\snnT{5.02}$.
The solid red curves in the plots correspond to mid-rapidity calculations and are compared with experimental data.
Good agreement with \cms\ data for both $\PbPb$ collisions throughout the whole $\pt$ range is observed.
For the lowest energy collision of $\AuAu$ at $\snnG{200}$ our results show consistency with data from the \STAR\ experiment for $\pt \geq \SI{4}{\GeV}$.
At the lower $\pt$ regime a complex interplay of different physical processes, is expected to occur.
One such physical process that is very important is the recombination mechanism which dominates the heavy quark hadronization at this regime and has been shown to decrease the $\raa$ at low $\pt$ while also increasing it at high $\pt$.
We also present predictions for forward rapidity $\raa$.

The calculations for $\Bmeson$ meson $\raa$ are shown in Fig.~\ref{fig:raaB}.
The same trend as the case for $\Dmeson$ mesons is observed at high $\pt$, where a larger parton suppression is observed in larger rapidity bins.

\begin{figure}[h!]
    \includegraphics{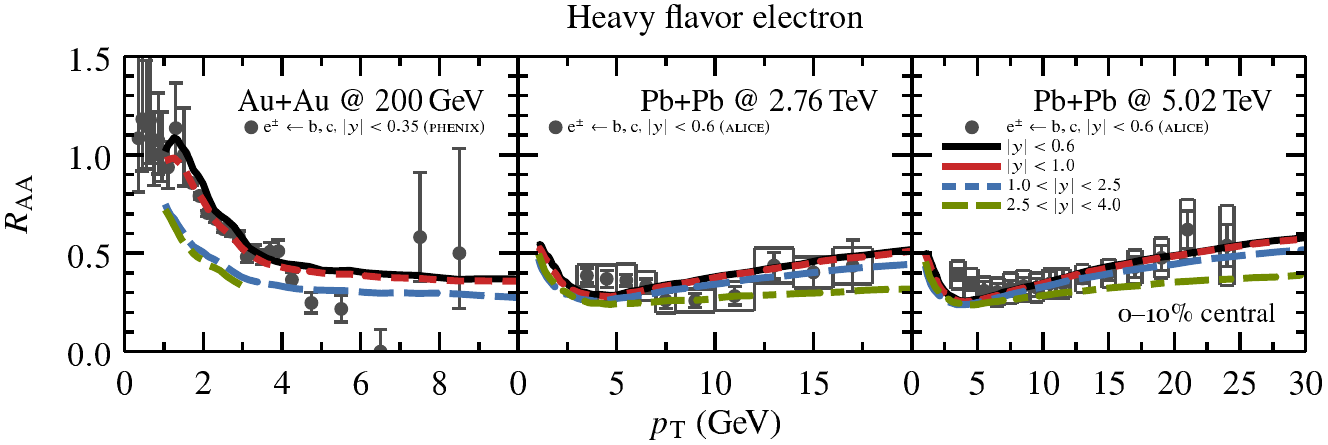}
    \caption{(Color online) Nuclear modification factor of heavy flavor electrons for central collisions in different ranges of rapidity.  Mid-rapidity experimental data from \phenix~\cite{Adare:2010de} (left), \alice~\cite{Adam:2016khe} (middle), and \alice~\cite{ALI-PREL-133360} (right) are also shown.}
    \label{fig:raae}
\end{figure}

\begin{figure}[b!]
    \includegraphics{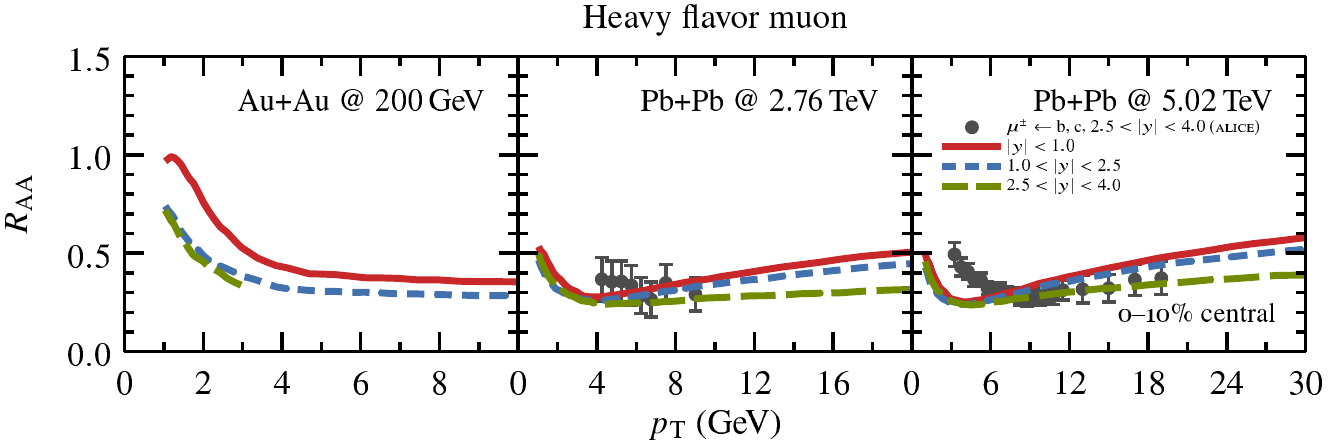}
    \caption{(Color online) Nuclear modification factor of heavy flavor muons for central collisions in different ranges of rapidity for $\AuAu$ at $\snnG{200}$ (left), $\PbPb$ at $\snnT{2.76}$ (middle) and $\PbPb$ at $\snnT{5.02}$ (right).  Experimental data from \alice~\cite{Abelev:2012qh,ALI-PREL-133394} at forward rapidity is also shown.}
    \label{fig:raam}
\end{figure}

In Fig.~\ref{fig:raae} we show heavy flavor electron results.
In these plots the solid black curves with a different rapidity range than that of the previous plots is used to compare our calculations results with experimental data from \alice\ at $\abs{y} < 0.6$.
For $\AuAu$ collisions we observe a good agreement with the \phenix\ data, despite the difference in the rapidity range, as it would be expected in this regime.
We also observe a good agreement for the $\PbPb$ collisions although our results slightly overestimate the data at $\snnT{2.76}$.

The results for the heavy flavor decayed muons are shown in Fig.~\ref{fig:raam}.
The comparison with experimental data shows good agreement especially for the larger beam energy at high $\pt$.
In these plots, however, since experimental data only spreads over a very limited range of $\pt$, it is harder to use data to discriminate between the different curves due to the proximity of the latter in this $\pt$ range.

Overall the above results show that when increasing the rapidity, we observe a larger suppression at the high $\pt$ regime, even though the expected medium size in these conditions is smaller.
Since $\raa$ not only depends on the path length experienced by the parton inside the medium, but also on the initial production spectra, these two effects compete with each other in the final result.
Here, a stronger effect from the initial heavy quark spectra, which is steeper for large rapidities in comparison with mid-rapidity, is observed to dominate in this region of $\pt$ leading to a larger suppression of the $\raa$.

\section{Conclusions}

We present the longitudinal dependence of heavy flavor nuclear modification factor obtained by coupling the (3+1)-dimensional viscous hydrodynamic medium background modeled by \clvisc\ with a relativistic Langevin equation based transport model incorporating both collisional and radiative energy loss.
The observed results are consistent with currently available experimental data and predictions for forward rapidity $\raa$ of heavy flavor mesons and leptons are presented for three different collision energies.
Our calculations have shown to indicate that at large rapidity, the smaller size of the medium and the steeper initial spectra of heavy quarks compete with each other in the resulting $\raa$.

Further studies on the longitudinal dependence of heavy flavor observables are needed to provide more sensitive constraints on the phenomenological models of the \qgp\ dynamics.

\section{Acknowledgments}

This work is supported in part by the Natural Science Foundation of China (\nsfc) under Grant Nos. \texttt{11775095}, \texttt{11890711}, \texttt{11890714} and \texttt{11935007}, by the China Scholarship Council (\CSC) under Grant No. \texttt{201906775042}, by the U.S. Department of Energy (\doe) under grant  Nos. \texttt{DE-AC02-05CH11231} and \texttt{DE-SC0013460}, and by the U.S. Natural Science Foundation (\nsf) under grant Nos. \texttt{ACI-1550300} and No. \texttt{ACI-1550228} within the \jetscape\ Collaboration.

\bibliography{library}
\end{document}